\documentclass[
 reprint,
 amsmath,amssymb,
 aps,
 prx,
 pdf,
 groupedaddress,
]{revtex4-2}

\usepackage{graphicx}   % Essential for figures
\usepackage{dcolumn}    % Align table columns on decimal point
\usepackage{bm}         % Bold math symbols
\usepackage{physics}    % Simplifies derivative notation, etc.
\usepackage{siunitx}    % For proper SI units
\usepackage{hyperref}   % Hyperlinks (remove if needed for APS submission)
\usepackage{xcolor}     % Text coloring (use sparingly)
\usepackage{scalerel}
\usepackage{tikz}

\usetikzlibrary{svg.path}

\newcommand{\edd}{\eta_\text{dd}}

\definecolor{orcidlogocol}{HTML}{A6CE39}
\tikzset{
  orcidlogo/.pic={
    \fill[orcidlogocol] svg{M256,128c0,70.7-57.3,128-128,128C57.3,256,0,198.7,0,128C0,57.3,57.3,0,128,0C198.7,0,256,57.3,256,128z};
    \fill[white] svg{M86.3,186.2H70.9V79.1h15.4v48.4V186.2z}
                 svg{M108.9,79.1h41.6c39.6,0,57,28.3,57,53.6c0,27.5-21.5,53.6-56.8,53.6h-41.8V79.1z M124.3,172.4h24.5c34.9,0,42.9-26.5,42.9-39.7c0-21.5-13.7-39.7-43.7-39.7h-23.7V172.4z}
                 svg{M88.7,56.8c0,5.5-4.5,10.1-10.1,10.1c-5.6,0-10.1-4.6-10.1-10.1c0-5.6,4.5-10.1,10.1-10.1C84.2,46.7,88.7,51.3,88.7,56.8z};
  }
}

\newcommand\orcidicon[1]{\href{https://orcid.org/#1}{\mbox{\scalerel*{
\begin{tikzpicture}[yscale=-1,transform shape]
\pic{orcidlogo};
\end{tikzpicture}
}{|}}}}

\begin{document}

\title{Exploring molecular supersolidity via exact and mean-field theories:\\single-microwave shielding}

%Alternatives?
%Quantum Monte Carlo–Informed Mean-Field Theory for Supersolid Phases of Ultracold Microwave-Shielded Molecules
%Mean-Field Theory for Supersolid Molecular Phases Guided by Quantum Monte Carlo
%Supersolidity in Microwave-Shielded Molecules from a Quantum Monte Carlo–Based Mean-Field Approach
%Quantum Monte Carlo–Derived Mean-Field Model for Supersolid Phases of Ultracold Molecules

\author{Tiziano Arnone Cardinale\,\orcidicon{0009-0000-7916-5091}}
\email{tiziano.arnone\_cardinale@fysik.lu.se}
\author{Thomas Bland\,\orcidicon{0000-0001-9852-0183}}
\author{Stephanie M. Reimann\,\orcidicon{0000-0003-1869-9799}}
\affiliation{Mathematical Physics and NanoLund, LTH, Lund University, Box 118, 22100 Lund, Sweden}

\date{\today}

\begin{abstract}
Ultracold polar molecular gases, stabilized via microwave shielding, offer a compelling new platform to explore quantum many-body physics with strong, long-range, and anisotropic interactions. In this work, we develop an extended Gross-Pitaevskii approach to describe bosonic dipolar molecules under single-microwave shielding, incorporating their effective interactions and adapting the quantum fluctuation corrections accordingly. We benchmark this new beyond-mean-field description against exact path-integral Quantum Monte Carlo simulations. Focusing on the experimentally relevant regime of positive scattering lengths, we find excellent agreement across a range of quantum phases, including superfluid, supersolid, and droplet states. We show that elliptic microwave polarization induces fully anisotropic superfluidity, characterized by direction-dependent sound velocities along each spatial axis---an effect absent in atomic dipolar gases. To study collective excitations in confined geometries, we develop a quasi-one-dimensional theory that captures the emergence of roton softening, and demonstrate that roton instabilities can be driven solely by tuning the ellipticity. These predictions align with recent experimental observations. Furthermore, we find that the nature of the superfluid-to-supersolid transition is strongly influenced by the ellipticity, appearing sharp at low values and continuous at higher ones. This tunability offers a potential route for low-entropy preparation of molecular supersolids via adiabatic ramps. While current experiments often rely on double-microwave shielding to mitigate losses, we find that single-shielded molecules already exhibit rich and controllable many-body behavior with our framework readily extendable to the double-shielded platform. This work establishes a versatile theoretical foundation for ultracold molecular gases and paves the way for future studies involving more complex shielding schemes.

\end{abstract}

\maketitle

\section{Introduction}
The advent of ultracold polar molecular gases has opened a powerful new avenue for exploring complex quantum many-body phenomena, enabled by their rich internal structure and strong long-range, anisotropic interactions. This opens exciting opportunities for new applications in quantum computation~\cite{demille2002quantum,albert2020robust,cornish2024quantum} and  simulation~\cite{micheli2006toolbox,baranov2012condensed,altman2021quantum}, sensing  
and metrology~\cite{flambaum2007enhanced,hudson2011improved,hutzler2020polyatomic}, or  
ultra-cold chemistry~\cite{krems2008cold,liu2022bimolecular,karman2024ultracold}. Rapid progress with 
 controlling and cooling molecular gases, as reviewed in~\cite{langen2024quantum}, gave access to entirely new concepts for quantum state engineering. Cold molecular ions~\cite{deiss2024cold} are another example, offering favorable electric field sensitivities for metrology and quantum information.

Experimentally, tremendous progress has been made in reaching the quantum-degenerate regime with ultracold molecules. Fermi gases of KRb and NaK have been cooled to degeneracy~\cite{de2019degenerate,valtolina2020dipolar,schindewolf2022evaporation,duda2023transition}, and recently the first Bose-Einstein condensates (BECs) of bosonic polar molecules were realized~\cite{bigagli2024observation,shi2025bose,Chomaz_NewsViews2024}, including evidence for droplet formation in NaCs~\cite{zhang2025observation}. These achievements depended critically on the development of collisionally protective techniques such as static electric field shielding~\cite{gorshkov2008suppression,avdeenkov2006suppression,wang2015tuning,gonzalez2017adimensional,matsuda2020resonant,li2021tuning,mukherjee2023shielding,mukherjee2024controlling} and microwave shielding~\cite{cooper2009stable,huang2012field,karman2018microwave,lassabliere2018controlling,karman2019microwave,karman2020microwave,valtolina2020dipolar,anderegg2021observation,karman2022resonant,schindewolf2022evaporation,lin2023microwave,deng2023effective,bigagli2023collisionally,chen2023field,wang2024prospects,chen2024ultracold,li2025tunable,xu2025effective,karman2025double,deng2025two,dutta2025universality}, suppressing inelastic losses while inducing long-range interactions. These techniques can be flexibly engineered through elliptic polarization~\cite{karman2019microwave,karman2020microwave,schindewolf2022evaporation,deng2023effective,chen2023field,chen2024ultracold,xu2025effective,zhang2025observation}, or extended to higher-order schemes such as double-microwave shielding~\cite{karman2025double,deng2025two,yuan2025extreme,bigagli2024observation,shi2025bose}.

Alongside these advances, a number of recent theoretical works have explored the ground-state phases and collective behavior of polar molecular systems. A Jastrow-variational approach~\cite{jin2025bose} and quantum Monte Carlo (QMC) simulations~\cite{zhang2025supersolid,langen2025dipolar,zhang2025quantum,ciardi2025self} were employed to incorporate the relatively strong correlations between polar molecules.  
These studies have confirmed the existence of  self-bound phases~\cite{jin2025bose}, revealed supersolid and crystalline droplet phases~\cite{zhang2025supersolid,ciardi2025self}, identified finite-temperature melting transitions~\cite{zhang2025quantum}, and demonstrated spontaneous pattern formation even without external trapping~\cite{ciardi2025self}. A common conclusion in this literature, however, is that mean-field theory is inadequate for describing these phases, particularly in regimes with strong correlations or negative effective scattering lengths~\cite{jin2025bose,langen2025dipolar}. Yet, many current experiments operate at positive scattering lengths, which is precisely the regime where a properly extended mean-field theory might apply, as hinted through a finite temperature study of ultra-cold molecules under double-microwave shielding with a circularly polarized field \cite{sanchez2025thermal}.

The present scenario historically mirrors the earlier development of dipolar atomic gas theory. The first dipolar Bose-Einstein condensate, realized in chromium~\cite{griesmaier2005bose}, showed excellent agreement with the dipolar Gross-Pitaevskii approach~\cite{goral2000bose,lahaye2009physics} in describing phenomena such as magnetostriction~\cite{stuhler2007magnetostriction} and anisotropic expansion~\cite{stuhler2005observation,giovanazzi2006expansion}. With the advent of more strongly dipolar species such as erbium and dysprosium~\cite{lu2011strongly,aikawa2012bose}, however, new effects emerged. In particular, the discovery of self-bound droplet phases in dysprosium~\cite{kadau2016observing}---inaccessible to standard mean-field theory---motivated the inclusion of the Lee-Huang-Yang (LHY) quantum fluctuation correction~\cite{lima2011quantum}, with similar self-bound states discussed in single component systems \cite{bulgac2002} and found in binary Bose gases~\cite{Petrov2015,PetrovAstrakharchik2016,Semeghini2018,Cabrera2018}. 
The extended Gross-Pitaevskii equation (eGPE), now widely used for dipolar systems~\cite{chomaz2016quantum,ferrier2016observation,wachtler2016quantum,bisset2016ground}, captures the essential features of dipolar superfluids and supersolids, as discussed in detail in the review by Chomaz {\it et al.}~\cite{chomaz2022dipolar}.

Intrigued by these parallels, we here revisit the applicability of mean-field theory to molecular systems, and develop the extended Gross-Pitaevskii equation for molecules (eGPE-M) incorporating the LHY correction specifically tailored to single-microwave-shielded interactions. In contrast to earlier claims of its failure~\cite{langen2025dipolar,boudjemaa2025nonequilibrium}, we demonstrate that when used within its regime of validity, i.e., positive scattering lengths, the eGPE-M agrees {\it quantitatively} with exact path integral Monte Carlo (PIMC) simulations across superfluid, supersolid, and single-droplet regimes.

\begin{figure}
    \centering
    \includegraphics[width=1\linewidth]{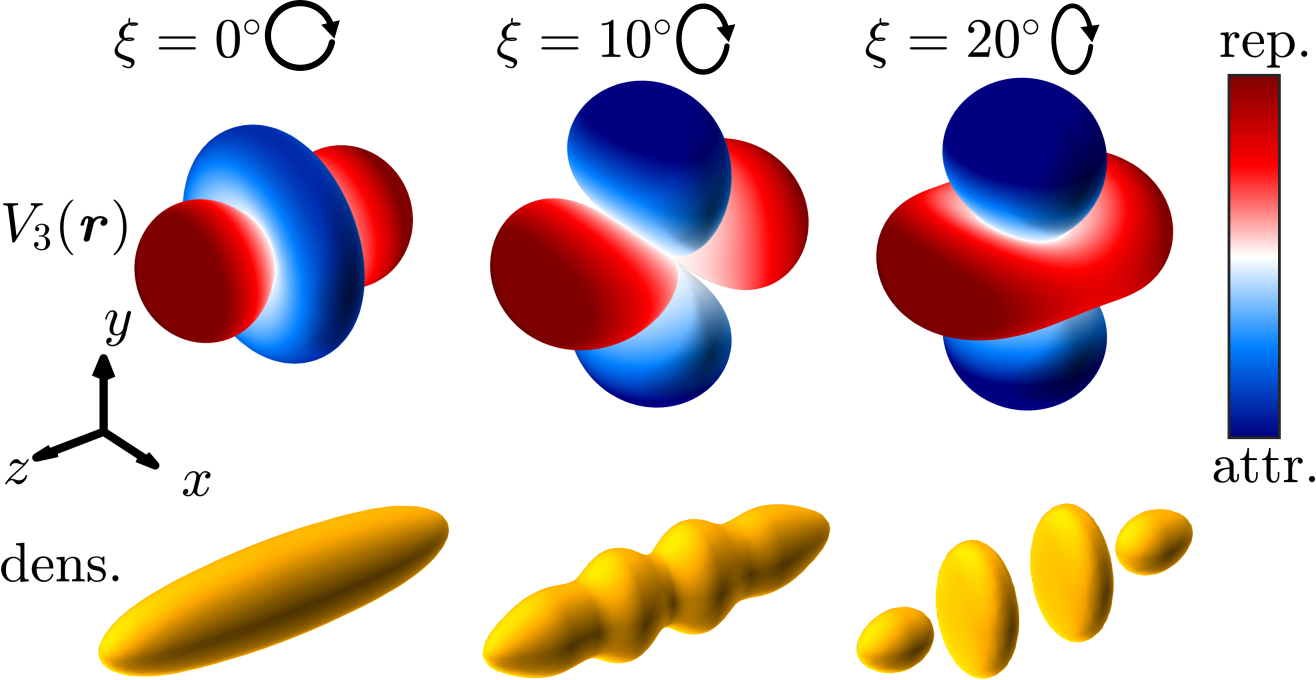}
    \caption{Spontaneous density modulation through microwave-shielding induced interaction shaping. Top row: Long-range interaction potential $V_3(\bm{r})$ in three-dimensional space, for increasing ellipticity $\xi$. Bottom row: Schematic density distributions of ultracold molecules in a cigar-shaped trap, showing a superfluid, supersolid, and independent droplet phase, induced through increasing ellipticity.}
    \label{fig:schem}
\end{figure}

We then use the eGPE-M to explore physical regimes uniquely accessible to single-microwave-shielded molecules. We show that elliptic polarization breaks cylindrical symmetry and induces fully anisotropic superfluidity, with distinct sound speeds along all three spatial directions. Through development of a quasi-one-dimensional theory, we show that roton excitations soften with increasing ellipticity and drive crystallization in cigar-shaped traps, consistent with recent QMC and experimental results~\cite{zhang2025observation,zhang2025supersolid,shi2025bose}. Moreover, we find that the nature of the superfluid-to-supersolid transition depends sensitively on the ellipticity parameter: it is sharp at low ellipticity and transforms into a continuous crossover at higher values. This tunability may enable low-entropy preparation of supersolids by crossing second-order transitions.

A schematic overview of the density modulations obtained for increasing ellipticity is shown in Fig.~\ref{fig:schem}. By comparing to PIMC simulations, the eGPE-M developed in this work is placed on firm footing as a new and intuitive, efficient tool to describe dipolar molecular gases, highlighting its limits and opportunities. 

\section{Two-body interaction potentials}
\subsection{Microwave shielding-induced interactions}\label{sec:V}
%The system we consider in this work is a gas of bosonic NaCs molecules, which are characterized by the dipole moment $d=4.6$ Debye and mass $m=155.8$ amu \cite{dagdigian1972molecular,aymar2005calculation}.
In order to realize the shielding potential, the bosonic molecules with mass $m$ are irradiated with a coherent microwave field that couples the rovibrational ground state with the rotational  $J=1$ manifold. The coupling is determined by the Rabi frequency $\Omega$, the detuning $\Delta$, and the ellipticity $\xi$ of the microwave field. The effective interaction potential between two molecules under single-microwave shielding (SMS) in the highest dressed-state channel is given by $V_{\mathrm{SMS}}(\bm{r}) = V_6(\bm{r}) + V_3(\bm{r})$~\cite{deng2023effective}. Here, in spherical coordinates,
\begin{align}\label{eqn:srpotential}
    V_6(\bm{r}) = \frac{C_6}{r^6}\sin^2 \theta&\big[1-\sin^2{2\xi}\cos^2{2\phi} \nonumber\\
    &+(1-\sin{2\xi}\cos{2\phi})^2\cos^2\theta\big]
\end{align}
where $C_6 = d^4 / (128\pi^2\epsilon_0^2\hbar\Omega(1+\delta^2)^{3/2})$ is the coefficient of the short-range shielding core potential, with dipole moment $d$, $\delta=|\Delta|/\Omega$, and
\begin{align}\label{eqn:lrpotential}
    V_3(\bm{r}) =
    \frac{C_3}{r^3}\left(\frac{3\cos^2\theta-1}{2}+ \frac{3}{2}\sin{2\xi}\cos{2\phi}\sin^2\theta\right)\,,
\end{align}
where $C_3 = d^2 / (24\pi\epsilon_0(1+\delta^2))$ is the coefficient of the long range dipole-dipole interaction. Note, that previous studies---for example, Refs.~\cite{karman2022resonant,bigagli2023collisionally,lin2023microwave,deng2023effective,wang2024prospects,langen2025dipolar,xu2025effective,karman2025double,dutta2025universality,ciardi2025self}---have a slightly different definition of $C_3$, choosing to absorb the factor of 1/2 found inside the brackets (see discussion on this choice below). It has been previously discussed that this interaction form is valid only for $\xi\lesssim 15^\circ$~\cite{deng2023effective}. However, the expression remains valid without any approximation in $\xi$, as we show in Appendix \ref{app:effpot}.
%Previous experimental works~\cite{bigagli2023collisionally} have shown that for NaCs molecules the three-body losses are lowest for $\Omega/2\pi$ between 4 MHz and 10 MHz. Therefore, we fix $\Omega = 2\pi\times 4\,$MHz throughout this work.

We begin by investigating the long-range interaction potential Eq.~\eqref{eqn:lrpotential} in more detail. Taking the direction of the wave vector along the $z$-axis, the potential in Cartesian coordinates takes the form
\begin{align}
    V_3(\bm{r}) = \frac{C_3}{r^3}\left[\frac{1}{2}\left(3\frac{z^2}{r^2} - 1\right) + \frac32\sin2\xi\frac{x^2 - y^2}{r^2}\right]\,.
\end{align}
Adding and subtracting $y^2+z^2$ from the numerator of the final term, this can be trivially written as 
\begin{align}
    V_3(\bm{r}) = \frac{C_3}{r^3}\Bigg((1-\sin2\xi)&\frac12\left(3\frac{z^2}{r^2}-1\right)\nonumber\\+\sin2\xi&\left(1-3\frac{y^2}{r^2}\right)\Bigg)\,,
    \label{eqn:realspace_rearranged_cart}
\end{align}
which returning to spherical polar coordinates is
\begin{align}
    V_3(\bm{r}) = \frac{C_3}{r^3}\big[&(1 - \sin2\xi)\frac{3\cos^2\theta - 1}{2} \nonumber \\&+ \sin2\xi(1 - 3\sin^2\theta\sin^2\phi)\big]\,.
    \label{eqn:realspace_rearranged}
\end{align}
This form of the expression reveals several key features of the interaction. The first term in Eq.~\eqref{eqn:realspace_rearranged} is an \emph{antidipolar} interaction, meaning it is repulsive only along the direction of the wavevector. This term supports density modulation along the repulsive axis. In ultracold atomic systems with antidipolar interactions, such modulation is predicted to lead to circularly symmetric pancake-like stacks, with tunable interlayer connectivity governed by interactions, trap geometry, and atom number~\cite{mukherjee2023supersolid,kirkby2023spin,mukherjee2025selective}.

The second term in Eq.~\eqref{eqn:realspace_rearranged} corresponds to purely \emph{dipolar} interactions, polarized perpendicular to the first term. These interactions are repulsive orthogonal to the polarization and give rise to structures with up to two broken translational symmetries. In a system of ultracold dipolar atoms, such observed configurations include linear droplet chains~\cite{boettcher2019transient,tanzi2019observation,chomaz2019long}, zig-zag patterns~\cite{norcia2021two}, and triangular lattices~\cite{norcia2021two,bland2022two}. At higher densities, theory predicts a rich variety of exotic phases---many yet to be observed~\cite{zhang2019supersolidity,hertkorn2021pattern,zhang2021phases,ripley2023two,zhang2024metastable}. With molecules, they may become accessible thanks to the stronger long-range interactions~\cite{schmidt2022self}.

In this work, we assume that the direction of the wave vector is along the $z$ axis, as in Eq.\,\eqref{eqn:realspace_rearranged_cart}. We find that at $\xi=0$ the interaction is purely antidipolar along $z$, potentially supporting modulation along that direction. As $\xi$ increases toward $\pi/4$, the antidipolar contribution diminishes while dipolar interactions along $y$ dominate, enabling crystallization in the $xz$-plane. This behavior should be expected as $\xi = \pi/4$ corresponds to linear polarization of the microwave field. Thus, single-microwave shielding alone appears insufficient to enable fully three-dimensional supersolidity. A similar analysis shows that negative values of $\xi$ correspond to dipolar alignment along $x$.

Finally, we clarify our choice to remove a factor of $1/2$ from $C_3$ relative to earlier studies of single-microwave shielding. This ensures that the interaction potential $V_3(\bm{r})$ correctly describes dipolar and antidipolar interactions with their physical strength and range, and allows us to define $g_\text{dd}=4\pi\hbar^2 a_\text{dd}/m = 4\pi C_3/3$ with dipolar length $a_\text{dd}$ consistent with the definition for dipolar atomic systems \cite{lahaye2009physics}. Including the factor of $1/2$ in $C_3$, as adopted in several prior works~\cite{karman2022resonant,bigagli2023collisionally,lin2023microwave,deng2023effective,wang2024prospects,langen2025dipolar,xu2025effective,karman2025double,dutta2025universality,ciardi2025self}, effectively halves the interaction strength while doubling its spatial range, which does not change the underlying physics but alters the interpretation of the dipolar length $a_\text{dd}$ and its relation to van der Waals interactions~\footnote{Consistency may suggest rescaling $C_6$ by a factor of $1/4$. However, this has no effect on the scattering length, and we thus retain its conventional definition.}.
%In this work, $a_\text{dd}$ ranges from 140nm at $\delta=3$ down to 20nm with $\delta = 8$.
One may also introduce a length scale associated to the $V_6$ potential. However, it is more convenient to define $R_{\Omega}=\left[d^2/\epsilon_0m\Omega^2\right]^{1/5}$, which is $\delta$-independent. In this way, the physical properties of molecular systems in free space are uniquely determined by $a_\mathrm{dd}$ and $R_\Omega$, where the former depends on $\delta$ and the latter depends on $\Omega$. In Sec.~\ref{sec:exp}, we contextualize the dimensionless units used throughout this work by mapping them to experimentally relevant parameters for various molecular species.

\subsection{Pseudo-potential approximation}

A main goal of this work is to compare the results from PIMC with mean-field theory and show that a mean-field description of the molecular condensate is viable in experimentally relevant parameter regimes. However, the potential in Eq.~\eqref{eqn:srpotential} cannot be used directly in numerical mean-field simulations as its Fourier transform is not well defined~\footnote{Introducing a short-range cutoff cures this problem, but it becomes difficult to estimate corrections due to quantum fluctuations~\cite{deng2023effective,boudjemaa2025nonequilibrium}.}. We thus consider the pseudo-potential $V'_{\mathrm{SMS}}(\bm{r}) = V_6'(\bm{r})+V_3(\bm{r})$, where we have replaced the shielding term by the zero-range contact potential commonly used for ultracold atomic gases \footnote{Other choices of $V'_\mathrm{SMS}$ are viable. See for instance Ref.~\cite{karman2025double} for an example with double-microwave shielding.}, 
\begin{align}
    V_{6}'(\bm{r}) = g\,\delta(\bm{r})\,.
    \label{eqn:sr}
\end{align}
This should be a good approximation as long as the system is sufficiently dilute and the s-wave scattering length $a_s'$ of $V_\mathrm{SMS}'$ is the same as $a_s$ of the original molecular potential $V_\mathrm{SMS}$. To prove this point, we consider the scattering problem of two molecules in the center-of-mass frame and write the scattering amplitude as
\begin{align}
    f(\hat{\bm k}_i,\hat{\bm k}_o) = \frac{4\pi}{k} \sum_{\ell\ell'mm'}i^{\ell-\ell'}Y_{\ell}^m(\hat{\bm k}_i)^* T^{mm'}_{\ell\ell'}Y_{\ell'}^{m'}(\hat{\bm k}_o),
    \label{eqn:scatt}
\end{align}
where $\hat {\bm k}_i$ and $\hat{\bm k}_o$ are the ingoing and outgoing directions, respectively, $Y_\ell^m$ are orthonormal spherical harmonics, and $T^{mm'}_{\ell\ell'}$ is the $k$-dependent scattering $T$-matrix. In the zero-energy limit we further define the $t$-matrix as $t^{mm'}_{\ell\ell'}=\lim_{k\to 0}T^{mm'}_{\ell\ell'}/k$ and the s-wave scattering length as $a_s=-t^{00}_{00}$. The $t$-matrix elements for $V_{\mathrm{SMS}}(\bm r)$ are calculated numerically by solving the two-body scattering problem at low energy by means of the Johnson log-derivative method~\cite{johnson1973the}.

\begin{figure}
    \centering
    \includegraphics[width=\linewidth]{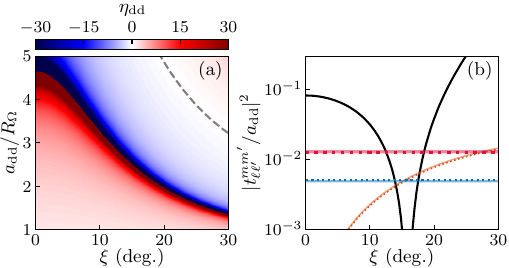}
    \caption{Efficacy of the pseudo-potential approximation. (a) s-wave scattering length $a_s$ as a function of $a_\mathrm{dd}$ and $\xi$. The grey-dashed line highlights the position of a field-linked resonance. (b) Plot of $t$-matrix amplitudes $t^{mm'}_{\ell\ell'}$ for $a_\mathrm{dd}=2.4 R_\Omega$: $t_{00}^{00}$ (black), $t_{02}^{00}$ (red), $t_{20}^{00}$ (blue) and $t_{02}^{02}$ (orange). Solid lines correspond to the original potential, while dotted lines correspond to the pseudo-potential $V_\text{SMS}'$. By construction, $t_{00}^{00}$ is the same for $V_\mathrm{SMS}$ and $V_\mathrm{SMS}'$.}
    \label{fig:as}
\end{figure}

In Fig.\,\ref{fig:as}(a) we show the calculated dimensionless interaction ratio $\eta_\mathrm{dd}=a_\mathrm{dd}/a_s$ as a function of $a_\mathrm{dd}/R_\Omega$, tunable through the detuning $\delta$, and $\xi$. The scattering length is positive for $a_\mathrm{dd}/R_\Omega\lesssim4.6$ at $\xi=0$. Increasing $\xi$, $a_s$ decreases due to the attractiveness of the long-range interaction and the reduction of the shielding core, and eventually becomes negative yielding ${\edd<0}$. For even larger $\xi$, a field-linked resonance appears \cite{avdeenkov2002collisional,avdeenkov2003linking,avdeenkov2004field,cooper2009stable,huang2012field,lassabliere2018controlling,chen2023field,chen2024ultracold}, corresponding to the formation of a shallow bound state.

In order to confidently employ Eq.\,\eqref{eqn:sr}, both $V_{\mathrm{SMS}}$ and $V'_{\mathrm{SMS}}$ must give the same scattering behavior. To show this is the case, we consider the pseudopotential $V'_{\mathrm{SMS}}(\bm r)$ and calculate the corresponding $t$-matrix within the first Born approximation using the relation
\begin{align}
    f^{\text{Born}}(\hat{\bm k}_i,\hat{\bm k}_o)=-\frac{m}{4\pi\hbar^2}\tilde{V}'_{\mathrm{SMS}}({\bm k}_i-{\bm k}_o)
    \label{eqn:scatt2}
\end{align}
for the scattering amplitude at zero energy. Therefore, in order to have $a_s=a_s'$ we must set ${g=4\pi\hbar^2a_s/m}$. We compare the $t$-matrices obtained from Eqs.~\eqref{eqn:scatt} and~\eqref{eqn:scatt2}. In Fig.\,\ref{fig:as}(b) we show the $t$-matrix amplitudes for ${a_\mathrm{dd}=2.4 R_\Omega}$ for even values of $\ell$ and $\ell'$, which are the only contributing terms in bosonic scattering under a parity-conserving potential. A similar comparison for double-microwave shielding was presented in a recent work~\cite{xu2025effective} for $\xi=0^\circ$ and $\xi=5^\circ$. Importantly, we find remarkable agreement between the two potentials across the entire range of $\xi$-values considered in this work. Furthermore, the s-wave contribution (black dashed line) dominates over the other contributions for a broad range of $\xi$. We also note that for $\xi=0^\circ$, where one recovers the azimuthally symmetric antidipolar potential, we find $t_{\ell\ell'}^{mm'}=0$ for $m\neq m'$.

\section{Methods}

The underlying Hamiltonian for an $N$-molecule dipolar system under harmonic confinement and SMS is
\begin{align}
    \hat{H} = \sum_i\hat{h}_0^{(i)} + \sum_{i<j}V_{\mathrm{SMS}}(\hat{\bm{r}}_i - \hat{\bm{r}}_j)\,,
    \label{eqn:H}
\end{align}
where $\hat{h}_0^{(i)} = -\frac{\hbar^2}{2m} \nabla_i^2 + V_{\mathrm{trap}}(\hat{\bm{r}}_i)$ is the single-particle Hamiltonian and $V_{\mathrm{trap}}(\bm{r}) = \frac{m}{2}(\omega_x^2 x^2 + \omega_y^2 y^2 + \omega_z^2 z^2)$ defines the external confinement. The two-body interaction potential $V_{\mathrm{SMS}}(\bm{r})$ was described in Sec.~\ref{sec:V}. In this section, we introduce the two complementary theoretical approaches used to solve Eq.~\eqref{eqn:H}. The first is a molecular mean-field theory we derive specifically for single-microwave-shielded molecules. The second is the path integral Monte Carlo method, well established for ultracold molecular systems, which we use to benchmark the mean-field results.

\subsection{Extended Gross-Pitaevskii equation for molecules under single-microwave shielding (eGPE-M)}
As discussed in the introduction, ultracold molecules exhibit significantly stronger dipolar interactions than their atomic counterparts~\cite{bigagli2024observation,shi2025bose,zhang2025observation}, making it essential to include quantum fluctuations in any mean-field description. To derive the corresponding correction, one requires two ingredients: the effective interaction potential of the molecular dipole-dipole interaction in momentum space and the corresponding Lee–Huang–Yang (LHY) term.

We begin by evaluating the long-range potential in momentum space via the Fourier transform 
$\tilde{V}_3(\bm{k}) = \int \mathrm{d}^3\bm{r}\, V_3(\bm{r})e^{i\bm{k} \cdot \bm{r}}$, performed in spherical coordinates. Using the plane wave expansion
\begin{align}
    e^{i \bm{k} \cdot \bm{r}} = 4\pi \sum_{\ell=0}^{\infty} i^\ell j_\ell(kr) \sum_{m=-\ell}^{\ell} Y_\ell^m(\hat{\bm{k}})^* Y_\ell^m(\hat{\bm{r}})\,
\end{align}
with spherical Bessel functions $j_\ell(kr)$, and the identities
\begin{align}
    3\cos^2\theta - 1 &= \sqrt{\frac{16\pi}{5}}\,Y_2^0(\hat{\bm{r}})\,, \\
    \cos(2\phi)\sin^2\theta &= \frac{1}{6}\sqrt{\frac{96\pi}{5}}\left(Y_2^2(\hat{\bm{r}}) + Y_2^{-2}(\hat{\bm{r}})\right)\,,
\end{align}
the orthogonality of the spherical harmonics allows the angular integral to be evaluated directly, yielding
\begin{align}
    \tilde{V}_3(\bm{k}) = g_\text{dd}\left(\frac{1 - 3\cos^2\theta_k}{2} - \frac32\sin 2\xi \cos 2\phi_k \sin^2\theta_k\right)\,,
    \label{eqn:DDIk}
\end{align}
where $(\theta_k, \phi_k)$ are the spherical angles of $\bm{k}$. As a reminder, we have introduced $g_\text{dd}=4\pi C_3/3$. The total effective interaction in momentum space is then $\tilde{V}'_{\mathrm{SMS}}(\bm{k}) = g + \tilde{V}_3(\bm{k})$.

With this expression, we now derive the LHY correction applicable to ultracold molecular gases under single-microwave shielding with arbitrary polarization. The shift in ground-state energy of the homogeneous molecular gas due to quantum fluctuations, $\Delta E$, can be calculated using the standard LHY result~\cite{lee1957eigenvalues}
\begin{align}
    \frac{\Delta E}{\mathcal{V}} = &\frac{1}{2} \int \frac{\mathrm{d}^3\bm{k}}{(2\pi)^3} \Bigg[ 
    \sqrt{ \frac{\hbar^2 k^2}{2m} \left( \frac{\hbar^2 k^2}{2m} + 2n_0 \tilde{V}'_{\mathrm{SMS}}(\bm{k}) \right)} \nonumber \\
    & - \frac{\hbar^2 k^2}{2m} - n_0 \tilde{V}'_{\mathrm{SMS}}(\bm{k}) 
    + \frac{m \left(n_0 \tilde{V}'_{\mathrm{SMS}}(\bm{k})\right)^2}{\hbar^2 k^2} \Bigg]\,,
\end{align}
where $n_0$ is the homogeneous density with volume $\mathcal{V}$.

Evaluating this integral yields the result
\begin{align}
    \frac{\Delta E}{\mathcal{V}} = \frac{128}{15\sqrt{\pi}} \frac{\hbar^2}{m} (a_s n_0)^{5/2} \, \mathrm{Re}\left[\mathcal{K}_5(\edd, \xi)\right]\,,
\end{align}
where the dimensionless function $\mathcal{K}_n$ is defined as
\begin{align}
    \mathcal{K}_n(\edd, \xi) = &\int_0^{2\pi} \mathrm{d}\phi_k \int_0^1 \mathrm{d}x 
    \bigg[1 + \edd \bigg(\frac{1 - 3x^2}{2} \nonumber \\
    & - \frac32\sin 2\xi \cos 2\phi_k (1 - x^2)\bigg)\bigg]^{n/2}\,,
    \label{eqn:K5}
\end{align}
and with the dimensionless relative interaction ratio $\edd = g_\text{dd}/g$, noting $\mathcal{K}_5$ becomes complex for $\edd>1$. With our choice of $C_3$, the system is dominantly dipolar for $\edd > 1$ and dominantly contact-like for $\edd < 1$,  independent of $\xi$. Equation~(\ref{eqn:K5}) extends previous works~\cite{langen2025dipolar,sanchez2025thermal,boudjemaa2025nonequilibrium} on the beyond-mean-field equation for single-microwave-shielding by the inclusion of ellipticity. A simple expansion around $\edd = 0$ provides insight into the $(\edd, \xi)$-dependence:
\begin{align}
    \mathcal{K}_5(\edd, \xi) \approx 2\pi\left(1 + \frac{3}{2} \edd^2 \frac{1 + 3\sin^2 2\xi}{4}\right)\,,
    \label{eqn:K5app}
\end{align}
revealing that the beyond-mean-field correction grows with both $\edd$ {\it and} ellipticity. Since we primarily work in the regime $\edd \gg 1$, we compute Eq.~\eqref{eqn:K5} numerically throughout; see App.~\ref{app:K5} for a comparison between the analytic expansion and the full integral. Note that for $\xi=\pi/4$ one recovers the LHY correction for dipolar atoms, $\mathcal{Q}_5(x)\equiv\mathcal{K}_5(x,\pi/4)$ (see, e.g., Ref.~\cite{lima2011quantum}), in-keeping with the observation that the dipolar potential becomes a standard dipole-dipole interaction polarized transversely to the microwave field propagation axis.

The extended Gross–Pitaevskii equation for the mean-field wavefunction $\Psi(\bm{r})$ describing an ensemble of zero-temperature molecules under single-microwave shielding incorporating quantum fluctuations is thus given by
\begin{align}\label{eqn:gpe}
    i\hbar \frac{\partial \Psi}{\partial t} = \bigg[
    & -\frac{\hbar^2}{2m} \nabla^2 + V_{\mathrm{trap}}(\bm{r}) 
    + \eta_\mathrm{QF} |\Psi(\bm{r})|^3 \nonumber \\
    & + \int \mathrm{d}^3 \bm{r}'\, V'_{\mathrm{SMS}}(\bm{r} - \bm{r}') |\Psi(\bm{r}')|^2 
    \bigg] \Psi(\bm{r})\,,
\end{align}
where the beyond-mean-field term is obtained from the local density approximation via $\partial(\Delta E/\mathcal{V}) / \partial n_0$, with coefficient
\begin{align}
    \eta_\mathrm{QF} = \frac{64 \hbar^2}{3m\sqrt{\pi}} a_s^{5/2} \, \mathrm{Re}\left[\mathcal{K}_5(\edd, \xi)\right]\,.
\end{align}
Throughout this work, we refer to Eq.~\eqref{eqn:gpe} as an extended Gross-Pitaevskii equation for molecules (eGPE-M). In what follows, we investigate ground-state solutions of the eGPE-M using the usual approach of the imaginary time propagation technique with split-step-Fourier methods. The dipolar term is evaluated efficiently in momentum space with a spherical cut-off applied to the potential to suppress aliasing effects; see App.~\ref{app:V3k} for details.

\subsection{Path Integral Monte Carlo}

In order to benchmark our mean-field results, we employ the well-established path integral Monte Carlo (PIMC) technique, that has been used in many settings for ultracold molecules under microwave shielding \cite{zhang2025supersolid,langen2025dipolar,zhang2025quantum,ciardi2025self}, and that we briefly detail below.
In the canonical ensemble, the system at finite temperature $T$ is described by the partition function $Z=\mathrm{Tr}[e^{-\beta \hat H}]$, where $\beta=(k_BT)^{-1}$ and observables can be sampled by the PIMC method, which is based on a high-temperature expansion of $Z$ \cite{ceperley1995path}. In this work, we employ the Takahashi-Imada approximation~\cite{takahashi1984monte} for the partition function, which reads $Z\simeq \text{Tr}[e^{-\tau \hat H_0}e^{-\tau \hat H'_1}]^M$, where $\tau=\beta/M$, $\hat H_0=\sum_i\hat h_0^{(i)}$ , $\hat U=\sum_{i<j}V_\mathrm{SMS}(\hat{\bm r}_i-\hat{\bm r}_j)$ and
\begin{align}
    \hat H'_1 &= \hat U+\frac{\tau^2}{24}[\hat U,[\hat H_0,\hat U]]
    = \hat U+\frac{\hbar^2\tau^2}{24m}\sum_{i=1}^N |\nabla_i \hat U|^2.
\end{align}
In the path-integral formalism, $Z$ becomes
\begin{align}
    Z &= \frac{1}{N!}\prod_{j=x,y,z}\left[\frac{1}{2\pi a_j^2\sinh(\hbar\omega_j\tau)}\right]^{\frac{NM}{2}} \nonumber \\ &\times\sum_{P}\int\prod_{\alpha=0}^{M-1}\prod_{i=1}^N \text{d}^3\bm r_{\alpha,i} e^{-S_0[\{\bm r_{\alpha,i}\}]-S_1[\{\bm r_{\alpha,i}\}]}, \label{eq:partition}
\end{align}
where $a_j$ is the harmonic oscillator length along $j$,  $\bm r_{\alpha,i}$ is the position of particle $i$ at bead $\alpha$, $M$ is the number of beads and $P$ is a permutation such that $\bm r_{M,i}=\bm r_{0,P(i)}$. Here, $S_0$ is the exact action of $N$ three-dimensional harmonic oscillators
\begin{align}
    S_0=\sum_{\alpha=0}^{M-1}\sum_{i=1}^N &\frac{(x_{\alpha,i}^2+x_{\alpha+1,i}^2)\cosh(\hbar\omega_x\tau)-2x_{\alpha,i} x_{\alpha+1,i}}{2\sinh(\hbar\omega_x\tau)a_{x}^2}\nonumber \\
    &+(x\to y)+(y\to z)
\end{align}
and the interacting action is
\begin{align}
    S_1 = \sum_{\alpha=0}^{M-1}\bigg(&\tau U(\bm r_{\alpha,1} ... \bm r_{\alpha,N})\nonumber
    \\&+\frac{\hbar^2\tau^3}{24m}\sum_{i=1}^N|\nabla_{\alpha,i} U(\bm r_{\alpha,1} ... \bm r_{\alpha,N})|^2\bigg)\,.
\end{align}
Monte Carlo updates are based on the staging algorithm and permutations are sampled by the worm algorithm~\cite{boninsegni2006worm}; see App.~\ref{app:V3k} for details on optimal parameter choices for our system.
%\newpage
\begin{figure}
    \centering
    \includegraphics[width=1\linewidth]{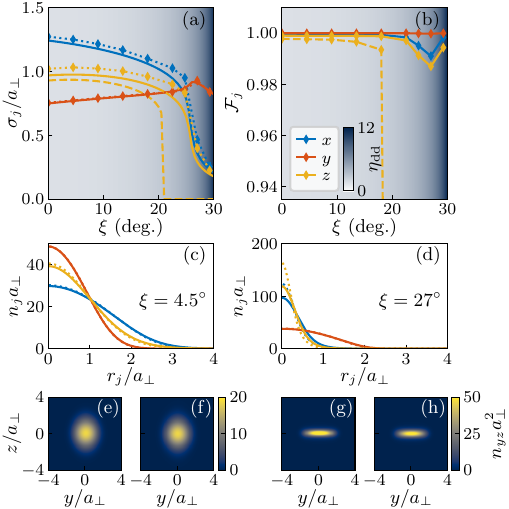}
    \caption{Agreement between mean-field theory and Monte Carlo methods in the superfluid-to-droplet regime, with $a_\mathrm{dd}=5.11\times10^{-2}a_\perp$. (a) Condensate widths as a function of ellipticity in the three cartesian directions. Solid lines correspond to eGPE-M, diamonds to PIMC. The dashed line shows $\sigma_z$ from the eGPE-M without the quantum fluctuation correction ($\eta_\text{QF}=0$), leading to collapse for $\xi>20^\circ$. (b) Fidelity, see Eq.\,\eqref{eqn:fid}. The background color of (a) and (b) corresponds to $\edd$. Panels (c) and (d) show linear densities for two representative values of $\xi$; colors and styles are the same as in (a) and (b). Column densities in the $yz$ plane corresponding to (c) are shown in (e) for eGPE-M and in (f) for PIMC; those corresponding to (d) are shown in (g) for eGPE-M and in (h) for PIMC. Peak densities for the highlighted states are (c) $5.03a^{-3}_\perp$ and (d) $90.88a^{-3}_\perp$.}
    \label{fig:superfluid}
\end{figure}

\section{Superfluid and droplet phase}\label{sec:super}
In order to explore the superfluid phase, we consider a gas of $N=100$ molecules. Inspired by the recent experiment of Ref.~\cite{bigagli2024observation}, where the trap frequencies were $\bm{\omega} = (40,80,40)\,$Hz, we take $\omega_x=\omega_z$ and $\omega_y=2\omega_x$. We define the transverse trap frequency $\omega_\perp=\sqrt{\omega_x\omega_y}$ and oscillator length $a_\perp=\sqrt{\hbar/m\omega_\perp}$, which we use as energy and length scales, respectively.

Our main goal in this section is to compare the eGPE-M and PIMC methods by selecting two representative values of the dipolar length $a_\mathrm{dd}$, which correspond to two different detunings $\delta$ for a fixed Rabi frequency $\Omega$, i.e. fixed $R_\Omega / a_\perp$. Specifically, we take $R_\Omega = 4.1 \times 10^{-2} a_\perp$, and explore two values of the dipolar length: $a_\mathrm{dd} = 5.11 \times 10^{-2} a_\perp$ and $a_\mathrm{dd} = 9.82 \times 10^{-2} a_\perp$ (see Sec.~\ref{sec:exp} for physical parameter ranges). For each case, we simulate the range of ellipticities $\xi$ where the effective scattering length satisfies $a_s > 0$. The eGPE-M results are obtained by solving Eq.~\eqref{eqn:gpe} in imaginary time, while the temperature for PIMC is set to $k_B T = 0.92 \hbar \omega_\perp$. We have verified that smaller temperatures do not yield a different result but do increase the computational cost. From the two methods, we compare observables such as the integrated density profiles $n_j$, the profile widths $\sigma_j = \sqrt{\langle r_j^2\rangle}$, and measure the fidelity
\begin{align}
    \mathcal{F}_j = \frac1N\int \text{d}r_j\, \sqrt{n_j^\text{(QMC)}n_j^\text{(eGPE-M)}}
    \label{eqn:fid}
\end{align}
for each cartesian component $j$, and, for example, $r_x\equiv x$.

Figure~\ref{fig:superfluid} presents our first comparison between molecular Gross-Pitaevskii and PIMC methods. We begin with the case $a_\mathrm{dd}=5.11\times10^{-2}a_\perp = 1.24R_\Omega$, corresponding to a larger detuning, for which ellipticities up to $30^\circ$ are accessible. In Fig.~\ref{fig:superfluid}(a), we show the density widths $\sigma_{x,y,z}$ obtained from both the eGPE-M and PIMC, finding excellent agreement across the full range of $\xi$. Both methods predict a transition from a superfluid to a liquid droplet state around $\xi=27^\circ$, as evidenced by the rapid change in $\sigma_x$ and $\sigma_z$. They agree on the superfluid-to-droplet transition point to within $1^\circ$ of each other. The mean-field theory remains accurate even deep in the droplet regime, where the peak density reaches $n_\text{peak}a_\perp^3=90.88$ at $\xi=30^\circ$. This corresponds to a diluteness parameter of $n_\text{peak}a_\text{dd}^3\approx0.012$, significantly smaller than unity.  For comparison, we also show results from the bare mean-field theory, which omits quantum fluctuations. In this case, the predictions are noticeably less accurate in the superfluid regime. This incorrectly indicates a collapse at $\xi = 20^\circ$, unable to find the droplet solution.

The agreement between eGPE-M and PIMC is further supported by the fidelity metric $\mathcal{F}_j$ in Fig.~\ref{fig:superfluid}(b). The slight dip in $\mathcal{F}_j$ near the transition point (though it remains close to unity) reflects a small discrepancy in the predicted location of the phase boundary. Representative density profiles are shown in Figs.~\ref{fig:superfluid}(c)-(h), illustrating a low-density superfluid and a high-density droplet, with particularly striking agreement in the column density profiles.

\begin{figure}
    \centering
    \includegraphics[width=1\linewidth]{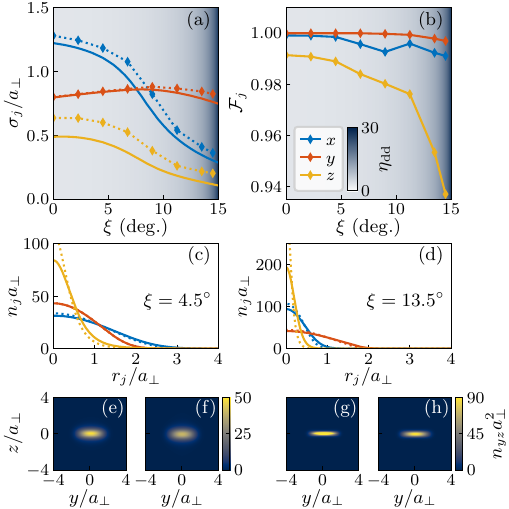}
    \caption{Agreement between mean-field theory and Monte Carlo methods in the superfluid-to-droplet regime, with $a_\mathrm{dd}=9.82\times10^{-2}a_\perp$. Note, there are no stable solutions when $\eta_\text{QF}=0$. Peak densities for highlighted states are (c) $18.42a^{-3}_\perp$ and (d) $171.9a^{-3}_\perp$. Panels are the same as in Fig.~\ref{fig:superfluid}.}
    \label{fig:superfluid2}
\end{figure}

We now turn to the case $a_\text{dd} = 9.82\times 10^{-2}a_\perp = 2.4R_\Omega$, corresponding to smaller detuning, where dipolar interactions are significantly stronger. Here, the region with positive $a_s$ is restricted to $\xi \in [0, 15]^\circ$. The corresponding results are shown in Fig.~\ref{fig:superfluid2}. In this regime, the eGPE-M without quantum fluctuations (setting $\eta_\text{QF}=0$) always leads to a collapse of the solution, highlighting the critical role of quantum fluctuations in stabilizing the gas. Despite the increased interaction strength---where $\edd$ spans from 3.5 to 30---the agreement between eGPE-M and PIMC remains remarkably good, and the condensate widths exhibit the same qualitative trend across ellipticity as found with $a_\text{dd} = 5.11\times 10^{-2}a_\perp$, and the fidelity $\mathcal{F}_j$ consistently exceeds 0.93. Even in the case of large $\xi = 13.5^\circ$ as shown in (d), the peak density is $n_\text{peak}a_\perp^3=171.9$ and the diluteness parameter is $n_\text{peak}a_\text{dd}^3 \approx 0.16$, still smaller than unity. Unlike the previous case, however, even for relatively small values such as $\xi = 4.5^\circ$, the condensate becomes notably elongated along the $y$-axis. As shown in Fig.~\ref{fig:superfluid2}(g) and (h), at $\xi = 13.5^\circ$ the system forms a narrow, cigar-shaped droplet. Finally, we note that at larger $a_\mathrm{dd}$ the state is always a liquid droplet, as has been discussed in previous theoretical works for circularly polarized SMS~\cite{langen2025dipolar,boudjemaa2025nonequilibrium}.

Our results in this section demonstrate a remarkable agreement between beyond-mean-field theory and exact QMC methods across both the superfluid and droplet regimes. The inclusion of quantum fluctuations via the LHY correction captures much of the essential physics, even in strongly dipolar scenarios. Nonetheless, minor deviations suggest that the LHY term alone may not always be sufficient for quantitative agreement. This is particularly evident in Fig.~\ref{fig:superfluid}(a) at $\xi=0^\circ$, where the LHY correction increases the density width towards the PIMC result, yet appears to underpredict the repulsive effect required for full agreement. Still, the level of consistency achieved is unprecedented: in the regime $\edd \gtrsim 1.5$, where stability is entirely provided by quantum fluctuations, our extended mean-field model remains highly accurate up to $\edd \approx 20$, and the sizable imaginary component of $\mathcal{K}_5$ does not appear to compromise the accuracy of the theory. This demonstrates that the extended mean-field framework is not only computationally efficient, but also sufficiently powerful to capture a broad spectrum of phenomena in dilute cold gases of dipolar molecules with positive scattering lengths.

\section{Excitation spectra}
Given the existence of stable single-droplet solutions, it is natural to ask whether this system can also support supersolid phases. To explore this possibility, we investigate the fate of linear excitations on a homogeneous background, seeking the emergence of finite-wavelength roton modes, widely regarded as precursors to supersolidity.

\begin{figure}
    \centering
    \includegraphics[width=1\linewidth]{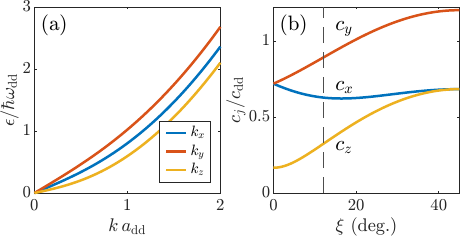}
    \caption{Fully anisotropic superfluidity in molecular Bose gases. (a) Excitation spectrum from Eq.~\eqref{eqn:exspec3D} with $n_0a_\text{dd}^3 = 0.2$, $\xi = 12^\circ$, and $\edd = 2$. Each curve $\epsilon (k_j)$ is evaluated along one axis (as indicated in the legend) with momenta in the orthogonal directions set to zero, i.e., $\epsilon(k_x) = \epsilon(k_x, 0, 0)$, etc.
(b) Directional speeds of sound $c_j$. The dashed line indicates the $\xi$ value used in (a). All other parameters are as in (a). We have defined the frequency scale $\omega_\text{dd} = \hbar/m a_\text{dd}^2$ and corresponding velocity scale $c_\text{dd} = a_\text{dd}\omega_\text{dd}$.}
    \label{fig:phonon}
\end{figure}

\subsection{Phonons and anisotropic superfluidity}
Following the standard theoretical treatment~\cite{pitaevskii2016bose,lima2012beyond}, we assume a homogeneous solution to Eq.~\eqref{eqn:gpe} as $n_0=|\Psi|^2$ and plane-wave excitations of the form $\delta\Psi\sim\exp(i\bm{k}\cdot\bm{r}-i\epsilon(\bm{k}) t/\hbar)$, to obtain the three-dimensional excitation spectrum satisfying
\begin{align}
    \epsilon^2(\bm{k}) = \frac{\hbar^2k^2}{2m} \left[ \frac{\hbar^2k^2}{2m} + 2n_0\tilde V'_{\mathrm{SMS}}(\bm{k}) + 3\eta_\text{QF} n_0^{3/2} \right]\,.
    \label{eqn:exspec3D}
\end{align}
Figure~\ref{fig:phonon}(a) shows the excitation spectrum for a representative case with intermediate ellipticity, $\xi = 12^\circ$, and dipolar strength $\edd = 2$. Each principal axis exhibits a distinct dispersion, indicating direction-dependent sound velocities and, consequently, fully anisotropic superfluidity---an effect beyond what is possible in cylindrically symmetric atomic dipolar gases. This prediction may prove valuable for future precision measurement technologies based on molecular superfluidity, where tunable, directionally dependent interactions could enable novel gyroscopic or accelerometer platforms. Note, that a phonon instability---unstable growth of imaginary modes at small momenta---occurs when $\edd\approx2.14$ for $\xi=0^\circ$ and increases to $\edd\approx18.42$ for $\xi=45^\circ$ due to the large increase in $\eta_\text{QF}$. In the absence of quantum fluctuations the instability instead occurs at $\edd=1$, independent of $\xi$.

To quantify the anisotropy, we extract the speed of sound along each spatial axis using
$c_j = \lim_{k_j \to 0} \epsilon(k_j)/\hbar k_j$,
where $\epsilon(k_j)$ denotes the excitation spectrum evaluated along the $j$th axis with momenta in the orthogonal directions set to zero. The resulting values are shown in Fig.~\ref{fig:phonon}(b). In dipolar systems, phonons propagating along the attractive axis incur a higher energy cost than those propagating perpendicularly~\cite{lahaye2009physics}, as density waves in the former must overcome attractive interactions between regions of high density, whereas in the latter they propagate without such resistance. As expected, in the limiting cases $\xi=0^\circ$ and $45^\circ$, the repulsive directions exhibit the lowest sound speeds. At all intermediate values of $\xi$, however, the breaking of azimuthal symmetry in Eq.~\eqref{eqn:realspace_rearranged} leads to distinct velocities along all three spatial directions.

We find no evidence of a roton minimum in the homogeneous molecular gas within this mean-field description, which is consistent with similar results for atomic dipolar gases \cite{lahaye2009physics}. However, recent theoretical work \cite{polterauer2025phases} suggests that strong correlations may induce one-dimensional density modulation even in the absence of a trap, though in a parameter regime beyond extended mean-field theory.

\subsection{Shielding-induced roton instabilities}

To probe spontaneous translational symmetry breaking in this system, it is necessary to introduce a trapping potential. It is known from dipolar atoms, for example, that a roton minimum typically emerges only when the system is confined along the direction where the interactions are attractive \cite{lahaye2009physics}. This confinement suppresses trivial head-to-tail alignment of the atoms and instead shifts collective excitations to finite momentum, enabling roton softening and the emergence of a crystalline order.

To explore excitations in the molecular regime, we derive a new quasi-one-dimensional approach, building upon previous work on dipolar atoms~\cite{blakie2020variational,pal2020excitations,blakie2020supersolidity}. We begin by separating the wavefunction into the form $\Psi(\bm{r}) = \psi_\text{1D}(z)\chi(x,y)$, where $\chi(x,y) = \exp\left[-(x^2/\ell_x^2 + y^2/\ell_y^2)/2\right]/\sqrt{\pi \ell_x \ell_y}$ is a normalized Gaussian with variational widths $(\ell_x, \ell_y)$. Substituting this Ansatz into the total energy functional and integrating over the transverse directions yields an effective quasi-one-dimensional energy functional, which we minimize to obtain the optimal variational widths:
\begin{align}
    E[\ell_x, \ell_y] &= \frac{\hbar^2}{4m} \left( \ell_x^2 + \ell_y^2 \right) 
    + \frac{n_\text{1D} g}{4\pi \ell_x \ell_y} \left(1 + \edd \tilde{V}^{\ell_x, \ell_y}_\text{1D}(0)\right) \nonumber \\
    &\quad + \frac{m}{4} \left( \omega_x^2 \ell_x^2 + \omega_y^2 \ell_y^2 \right)
    + \frac{4\eta_\text{QF}}{25 (\pi \ell_x \ell_y)^{3/2}} n_\text{1D}^{3/2}\,,
\end{align}
\begin{figure}
    \centering
    \includegraphics[width=1\linewidth]{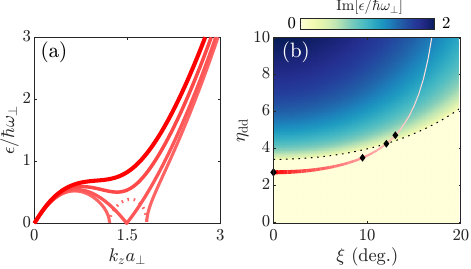}
    \caption{Roton instability of microwave-shielded molecules for fixed $a_\text{dd}  = 0.19a_\perp$, in an infinite tube with radial frequency $\omega_\perp$ and linear density $n_\text{1D}a_\perp = 18.6$. (a) Solutions to Eq.~\eqref{eqn:exspec1D} for ellipticities $\xi = (0,\,9.5,\,12.04,\,13)^\circ$, corresponding to $\edd = (2.72,\,3.51,\,4.269,\,4.73)$ using the calculated scattering lengths (Colors chosen to match the scattering lengths shown in Fig.~\ref{fig:as}). Dotted curve indicates the imaginary part of the excitation spectrum for $\xi=13^\circ$. (b) Phase diagram showing the imaginary component of the excitation spectrum as a function of $(\edd,\,\xi)$. The black dotted line marks the stability boundary separating stable and roton-unstable regimes. The red-to-pink curve traces the realistic $a_s$-dependent trajectory of varying $\xi$ alone, and diamonds indicate the parameter values plotted in (a).}
    \label{fig:roton}
\end{figure}
where $n_\text{1D} = |\psi_\text{1D}|^2$ is the homogeneous one-dimensional density. The approximate expression for the reduced dipole–dipole interaction, extending the results of Ref.~\cite{pal2020excitations}, is given by
\begin{align}
    \tilde{V}^{\ell_x, \ell_y}_\text{1D}(k_z) &= (1 - \sin 2\xi)\frac{3q_z e^{q_z} \text{Ei}[-q_z] + 1}{2} \nonumber \\
    &\quad + \frac{\sin 2\xi}{\ell_x + \ell_y} \left(3\ell_x q_y e^{q_y} \text{Ei}[-q_y] + 2\ell_x - \ell_y\right)\,,
\end{align}
where $\text{Ei}[x]$ denotes the exponential integral function, and
\begin{align}
    q_y = \frac{k_z^2}{2} \sqrt{\ell_y^3 \ell_x}\,, \quad 
    q_z = \frac{k_z^2}{2} \ell_x \ell_y \left( \frac{2 \ell_x \ell_y}{\ell_x^2 + \ell_y^2} \right)^{4/5}.
\end{align}
With these ingredients, the quasi-one-dimensional excitation spectrum becomes
\begin{align}
    \epsilon^2(k_z) = \frac{\hbar^2 k_z^2}{2m} \Bigg[ 
    &\frac{\hbar^2 k_z^2}{2m} 
    + \frac{n_\text{1D} g}{\pi \ell_x \ell_y} \left(1 + \edd \tilde{V}^{\ell_x, \ell_y}_\text{1D}(k_z)\right) \nonumber \\
    &+ \frac{6\eta_\text{QF}}{5 (\pi \ell_x \ell_y)^{3/2}} n_\text{1D}^{3/2} \Bigg]\,.
    \label{eqn:exspec1D}
\end{align}

\begin{figure*}[!t]
    \centering
    \includegraphics[width=\linewidth]{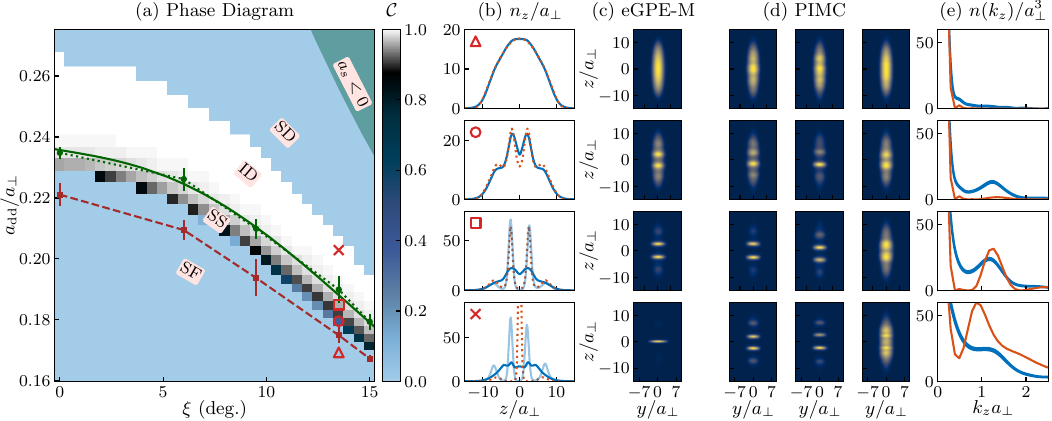}
    \caption{Supersolid phases of dipolar molecules in a cigar trap. (a) eGPE-M phase diagram based on the density contrast $\mathcal{C}$, highlighting superfluid (SF), supersolid (SS), independent droplet (ID), and single droplet (SD), and phonon unstable ($a_s<0$) states, as defined in the main text. The solid green curve denotes the SS-ID transition. Squares (bullets) indicate the SF-SS (SS-ID) transition points extracted from PIMC. Error bars correspond to half the spacing between neighboring values of $a_\mathrm{dd}/a_\perp$. (b) Axial linear density from PIMC (solid curves), showing a representative single run (light blue), the average over 48 runs (dark blue), and the corresponding eGPE-M result (red dotted curves). (c) Column densities in the $yz$-plane from eGPE-M, corresponding to states with $\mathcal{C} = (0, 0.37, 0.96, 1)$ from top to bottom. (d) Column densities from PIMC for single runs (left and center) and the averaged result (right). (e) Momentum distribution $n(\bm{k})$ along the $k_z$ direction, from PIMC (blue) and eGPE-M (red). PIMC data are averaged over all runs; the linewidth indicates statistical uncertainty.}
    \label{fig:SS}
\end{figure*}

Figure \ref{fig:roton} reveals the roton spectrum for the infinite tube scenario with radial frequency $\omega_\perp$. First, we show that for fixed $a_\text{dd}  = 0.19a_\perp=2R_\Omega$ and $\edd$ taken from Fig.~\ref{fig:as}(a), that increasing $\xi$ can lead to a roton instability. A similar observation has been made through double microwave-shielded gases by very recent PIMC calculations~\cite{zhang2025supersolid} and experimental observation of droplets~\cite{zhang2025observation} through varying ellipticity. Similar to dipolar atoms, the roton wavenumber is approximately $k_\perp=1.5/a_\perp\approx\sqrt{2}/a_\perp$ \cite{ronen2007radial,jonalasinio2013roton}. We note that this is only weakly dependent on ($\xi,\edd$). Finally, in Fig.~\ref{fig:as}(b), we show that the critical $\edd$ for instability depends on the ellipticity angle, but always $\edd>1$ as expected. By tuning both the detuning and ellipticity it should be possible to navigate this phase space, and map from an antidipolar roton instability to a dipolar one. Increasing the density lowers the critical $\edd$, though it never falls below 1.

\section{Supersolid phases of molecular Bose gases}\label{sec:SS}

We have shown that an infinite tube of molecules can undergo a roton instability. We now extend this analysis to a finite system by introducing an elongated trapping potential with aspect ratio $\omega_\perp/\omega_z=6$ and a larger number of molecules, $N = 250$. As in Sec.~\ref{sec:super}, we consider the case where the trap frequency $\omega_\perp$ and the Rabi frequency $\Omega$ are held constant and the detuning is varied to achieve different $a_\mathrm{dd}/a_\perp$ ratios. In this section, we consider $R_\Omega=9.44\times 10^{-2}a_\perp$. Within the eGPE-M formalism, we compute the ground-state phase diagram across a range of parameters where the dipolar length $a_\text{dd}$ is sufficiently large to support roton instabilities, as shown in Fig.~\ref{fig:SS}(a).

In this regime, we find that increasing either the ellipticity $\xi$ or the dipolar length $a_\mathrm{dd}$ leads to spontaneous translational symmetry breaking along the $z$-axis. To quantify the degree of crystallization, we use the density contrast $\mathcal{C} = (n_\text{max} - n_\text{min}) / (n_\text{max} + n_\text{min})$, where $n_\text{max}$ and $n_\text{min}$ are the local maximum and nearest minimum of the axial density profile, respectively. By this definition, $\mathcal{C} = 0$ corresponds to either a superfluid state or a single self-bound droplet. We identify states with $0 < \mathcal{C} \leq 0.98$ as supersolid~\footnote{The choice of $\mathcal{C}=0.98$ demarcating the transition from supersolid to independent droplet arrays is arbitrary, and studies on supersolidity often take a value between 0.95 and 0.99.}, characterized by partial crystallization and sufficient connectivity to support global phase coherence. In contrast, states with $0.98 < \mathcal{C} \leq 1$ correspond to isolated, phase-incoherent droplet arrays. Our phase diagram exhibits a broad range of supersolid states, over all $\xi$ and a range of $\delta$ that depends on the molecule and $a_\perp$ considered (species specific parameters are shown in Sec.~\ref{sec:exp}). These states are not all alike. At small $\xi\lesssim10^\circ$ the interaction is dominantly antidipolar, and the droplet aspect ratio in the $xy$-plane is close to unity. However, increasing $\xi$ enhances the dipolar contribution, eventually leading to an array of elongated droplets along $y$. Example densities are shown for $\xi = 13.5^\circ$ in panels (b) and (c), illustrating the progression from a smooth superfluid to a slightly modulated supersolid, a deeply modulated supersolid, and finally an isolated droplet state. Density isosurfaces of these three states are shown in Fig.~\ref{fig:schem}. The superfluid configuration exhibits a peak linear density of approximately $18.60/a_\perp$. From Fig.~\ref{fig:roton}, we predicted that an infinite tube at this density and $a_\text{dd}  = 0.19a_\perp$ would exhibit a roton instability at $\xi = 12.04^\circ$. Within the resolution of our phase diagram, the transition from superfluid-to-supersolid occurs between $12^\circ$ and $12.5^\circ$, demonstrating the predictive power of our reduced quasi-one-dimensional model in capturing the onset of crystallization.

We validate the eGPE-M phase diagram using PIMC simulations at $ k_BT = 0.49\hbar\omega_\perp$, evaluating the density contrast for multiple values of $\xi$. A subtlety arises when interpreting Monte Carlo results in the presence of spontaneous symmetry breaking. In the superfluid phase, different PIMC runs yield consistent density profiles. However, in the supersolid regime, each run may exhibit different configurations---varying in droplet number, size, or position---due to stochastic symmetry breaking. Averaging over these misaligned configurations can suppress contrast and obscure the underlying crystalline order. To address this, we perform 48 independent PIMC runs for each parameter set and compute the density contrast $\mathcal{C}$ individually for each realization.

We consider five values of $\xi$ and scan over $a_\mathrm{dd}$ to identify the boundaries between the superfluid, supersolid, and isolated droplet phases. Phase boundaries in Fig.~\ref{fig:SS}(a) are drawn using $\mathcal{C} > 0.35$ for the SF–SS transition and $\mathcal{C} > 0.98$ for the SS–ID crossover. Due to the finite simulation temperature and stochastic variation between runs, the contrast is never strictly zero; our choice of $0.35$ reflects the distinction between fluctuations in a superfluid and spontaneous density modulation indicative of supersolidity. The agreement between eGPE-M and PIMC is strong across all values of $\xi$. For the SF–SS boundary, the largest discrepancy is under 4\% in $a_\mathrm{dd}/a_\perp$, while the SS–ID crossover matches exactly. We attribute the slight deviation at the SF–SS boundary to finite-size effects in PIMC: this transition is a true phase transition, well-defined only in the thermodynamic limit. In contrast, the SS–ID transition is a smooth crossover, set by an arbitrary contrast threshold and therefore less sensitive to system size. Moreover, previous studies of dipolar atomic supersolids at finite temperature have shown that the transition shifts to weaker dipolar interaction strengths with increasing temperature~\cite{sanchez2023heating,bombin2025creating}. Thus, such discrepancies are to be expected when comparing to zero temperature theory.

Density distributions from PIMC are shown in Fig.~\ref{fig:SS}(b) and (d). In the supersolid regime, individual runs (light blue curves) clearly display the expected modulation, showing excellent agreement with the eGPE-M predictions (dotted lines). We note that the eGPE-M finds a three droplet stationary solution with only a relative energy change of 10$^{-8}$. The PIMC also finds three droplet solutions, hence the averaged profile becomes misleading. Unlike the eGPE-M, we do not observe single droplet states in the PIMC data for the values of $a_\mathrm{dd}$ shown. This discrepancy is likely due to the breakdown of the mean-field model at high densities. The supersolid state in Fig.~\ref{fig:SS}(b)$\square$ has a peak density of $n_{\text{peak}} = 8.24\, a_\perp^{-3}$, corresponding to a diluteness parameter $n_{\text{peak}} a_{\text{dd}}^3 \approx 0.05$, where both methods are in excellent agreement. However, the eGPE-M solution in Fig.~\ref{fig:SS}(b)$\times$ corresponds to a single droplet with peak density $n_{\text{peak}} = 12.14\, a_\perp^{-3}$, giving $n_{\text{peak}} a_{\text{dd}}^3 \approx 0.11$. We therefore conservatively estimate that the eGPE-M remains valid for $n_{\text{peak}} a_{\text{dd}}^3 \lesssim 0.075$.

While global phase coherence is implicit in the mean-field theory, it must be verified explicitly in QMC. To do so, we calculate the one-body density matrix (OBDM), $ \rho(\bm{r}, \bm{r}') = \langle \psi^\dagger(\bm{r}') \psi(\bm{r}) \rangle $, which captures the system's coherence. However, just like the density, the OBDM is affected by translational symmetry breaking in the SS regime.

To extract coherence in a translationally invariant way, we consider the momentum distribution,
\begin{align}
    n(\bm{k}) = (2\pi)^{-3} \int \! \text{d}^3\bm{r} \, \text{d}^3\bm{r}' \, e^{i \bm{k} \cdot (\bm{r} - \bm{r}')} \rho(\bm{r}, \bm{r}')\,,
\end{align}
which is the Fourier transform of $\rho(\bm r,\bm r')$ with respect to the relative coordinate $\bm r_\mathrm{rel} = \bm r-\bm r'$, and the centroid coordinate $\bm R= (\bm r+\bm r')/2$ has been integrated out. Hence, this is insensitive to global displacements of the cloud and provides a robust measure of long-range coherence.

Figure~\ref{fig:SS}(e) shows the momentum distribution averaged over 48 runs, alongside the corresponding eGPE-M prediction $ n(\bm{k}) = (2\pi)^{-3} |\tilde{\psi}(\bm{k})|^2 $. In the supersolid regime, the appearance of side peaks in both methods---centered around $k_z \sim 1.5/a_\perp$---confirms the presence of phase-coherent density modulations. This wavevector is in close agreement with the roton instability predicted by the reduced quasi-1D model in Fig.~\ref{fig:roton}. The match between PIMC and eGPE-M is excellent, except in the ID regime where the eGPE-M collapses into a single droplet.

% Paragraphs:
% \begin{itemize}
%     \item Everything PIMC, including contrast definition, temperature, setup, number of runs, any technical details. Here mention that the agreement of the boundary is great, and how the boundary is calculated, and that we think it is shifted up because of Fig.~\ref{fig:roton}, where the roton energy gap is comparable to the temperature about 1 deg. away from the transition point. (your points: discuss problem of metastable worldline configurations in PIMC (make a connection with experiments, explain definition of contrast for PIMC, Brag about agreement of the transition point!)
%     \item Discuss how the translational phase of the droplets becomes more random at smaller delta, hence the supersolidity gets washed out of the averaging. Hence, we introduce the momentum distribution stuff. Furthermore, given the definition for the momentum distribution for PIMC, this also proves global coherence, and supersolidity, compared to the GPE which is phase coherent by design.
%     \item The largest disagreement between the two methods occurs in the independent droplet regime, where the eGPE-M underestimates the droplet number and coherence. Here, the peak density is $n_\text{peak}=2.04\times10^{20}$m$^{-3}$, with a corresponding diluteness parameter of $n_\text{peak}a_\text{dd}^3\approx0.25$. From this, we conclude that our method works well for $n_\text{peak}a_\text{dd}^3\lesssim0.1$.
% \end{itemize}

\begin{figure}
    \centering    \includegraphics[width=1\linewidth]{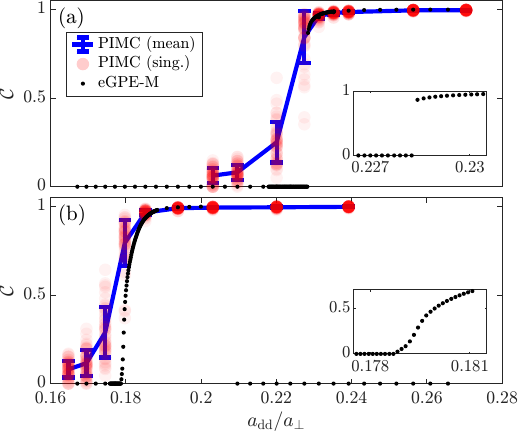}
    \caption{Interaction dependence of the superfluid-to-supersolid transition order. Ellipticity is fixed to (a)~$\xi = 0^\circ$ and (b)~$\xi = 13.5^\circ$. Each black dot shows the density contrast of the eGPE-M ground state. The average contrast extracted from PIMC (blue) is computed from 48 individual simulations (red points), with error bars indicating one standard error. Insets show a magnified view near the transition point. Other parameters are identical to those of Fig.~\ref{fig:SS}.}
    \label{fig:contrast}
\end{figure}

Our results uncover a striking feature of molecular supersolids not present in their atomic dipolar counterparts: the ability to tune the order of the superfluid-to-supersolid transition via the ellipticity of the microwave shielding. It is known that such a transition can be of second order when the roton wavelength on the superfluid side closely matches the modulation wavelength of the emerging supersolid phase~\cite{blakie2020supersolidity,smith2023supersolidity}, or when the transition is triggered by a roton immiscibility in two-component condensates~\cite{kirkby2023spin,kirkby2024excitations}, resulting in a continuous evolution of density contrast across the phase boundary. In Fig.~\ref{fig:contrast}, we demonstrate that this condition can be achieved or violated by varying the ellipticity parameter $\xi$. Panel (a) shows the contrast across the transition for the purely antidipolar case $\xi = 0^\circ$, where the system exhibits a sharp jump from an unmodulated superfluid to a strongly modulated density profile---indicative of a first-order transition. In contrast, for $\xi = 13.5^\circ$ [panel (b)], the contrast grows smoothly from zero, consistent with a continuous (second-order) transition, though a weakly first-order character cannot be fully excluded.

While such distinctions in transition order are difficult to resolve in a finite-sized system---due to enhanced fluctuations at a critical point, as seen in our PIMC simulations---the underlying difference remains physically meaningful. Importantly, continuous transitions, such as the one at larger $\xi$, are far more favorable for the adiabatic preparation of supersolid phases, as they minimize heating and help preserve global phase coherence. These findings suggest that ellipticity not only tunes the interaction anisotropy but also offers a powerful handle for optimizing the dynamics of molecular supersolid formation.

\section{Experimental Outlook}\label{sec:exp}

While our results have been presented in dimensionless units, we now contextualize them using typical parameters relevant to current ultracold molecular experiments, and discuss the feasibility of realizing supersolids under single-microwave shielding.

The main limitations in molecular gas experiments are two-body inelastic collisions and three-body losses. Microwave shielding significantly suppresses the former, but its effectiveness depends sensitively on the Rabi frequency $\Omega$ and detuning $\delta$, with optimal parameters varying across molecular species~\cite{langen2024quantum}. For example, in NaCs molecules, two-body inelastic collisions are reduced by over three orders of magnitude for $\delta \sim 1$ and $\Omega/2\pi = 4\,$MHz~\cite{bigagli2023collisionally}, though this suppression weakens at larger detuning. In contrast, three-body loss rates are minimized at larger detunings, $\delta \gtrsim 2$~\cite{stevenson2024three}, creating a trade-off. For lighter species such as CaF, shielding becomes effective only at higher Rabi frequencies ($\Omega/2\pi \gtrsim 5\,$MHz), while lower values can actually enhance losses~\cite{anderegg2021observation}.

\begin{figure}
    \centering
    \includegraphics[width=\linewidth]{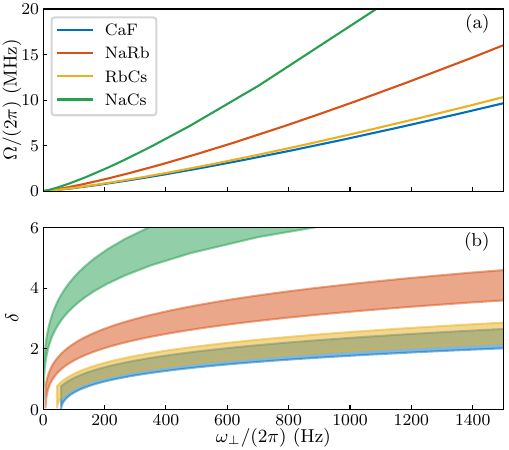}
    \caption{Experimental parameters required to produce Fig.~\ref{fig:SS}(a) for four exemplary molecular species. Rabi frequency (a) and detuning range (b) as a function of $\omega_\perp$. There is a minimum $\omega_\perp$ required to see crystallization set by the dipolar length at zero detuning, which is visible for CaF and RbCs.}
    \label{fig:CaF}
\end{figure}

Figure~\ref{fig:CaF} shows the parameter regimes corresponding to our chosen ratio $R_\Omega / a_\perp = 9.44 \times 10^{-2}$ for four different molecules studied experimentally~\cite{lin2023microwave,bigagli2023collisionally,anderegg2021observation,gregory2021}, plotted against the transverse trap frequency. From this figure, we can deduce that an ideal scenario may be reached with either CaF or RbCs with $\omega_\perp=2\pi\times1000\,$Hz, where $\Omega/2\pi \approx 5\,$MHz and $\delta\approx2$. We emphasize that tuning the ratio $R_\Omega / a_\perp$ would shift the accessible interaction strengths, enabling compatibility with different shielding conditions. Although no polar molecule has yet been brought to quantum degeneracy using SMS, recent work has demonstrated that the method is universally applicable across species~\cite{dutta2025universality}, suggesting that identifying favorable regimes with suppressed loss is a realistic near-term goal.

%https://journals.aps.org/pra/pdf/10.1103/PhysRevA.92.053401

% \begin{table}[h!]
% \centering
% \begin{tabular}{c|c|c|c|c}

%  & \textbf{NaRb} & \textbf{NaCs} & \textbf{CaF} & \textbf{RbCs}\\
% \hline\hline
% \boldmath$d$ \textbf{(Debye)} & 3.2 & 4.6 & 3.07 & 1.225 \\
% \boldmath $a_\mathrm{dd}|_{\delta=0}\,$\textbf{(\textmu{}m)} & 0.9326 & 2.735 & 0.4619 & 0.2739 \\
% \boldmath $\omega_\perp/(2\pi)$ \textbf{(Hz)} & 123 & 117 & 650 & 280\\
% \boldmath$\Omega / (2\pi)$ \textbf{(MHz)} & 0.705 & 1.235 & 3.388& 1.26 \\
% \boldmath$[\delta_\mathrm{min},\delta_\mathrm{max}]$ & [1.73,2.31] & [3.55,4.54] & [1.53,2.07] & [1.22,1.46] \\

% \end{tabular}
% \caption{Table with physical parameters. Here $\omega_\perp$ is chosen arbitrarily, $\Omega$ and $[\delta_\mathrm{min},\delta_\mathrm{max}]$ are the Rabi frequency and the detuning range needed to reproduce the phase diagram in Fig.~\ref{fig:SS}(a) for the particular choice of $\omega_\perp$.}
% \label{tab:exp}
% \end{table}

\section{Conclusion}

We have established a robust and predictive theoretical framework for ultracold molecular Bose gases under single-microwave shielding, bridging exact Quantum Monte Carlo simulations with an extended mean-field description that remains accurate across a wide range of experimentally relevant parameters. This agreement notably spans the superfluid, supersolid, and single-droplet regimes, validating the mean-field approach in capturing the rich phenomenology of microwave-shielded molecules. In the more strongly interacting limit, the theory eventually breaks down, and we tentatively identify a validity threshold at a diluteness parameter $n_\text{peak}a_\text{dd}^3\lesssim0.075$.

By analyzing the excitation spectrum of the homogeneous gas, we have demonstrated that elliptic microwave shielding leads to fully anisotropic superfluidity, where the speed of sound differs along all three spatial directions. This effect arises from azimuthal symmetry breaking in the effective interaction and represents a key departure from the cylindrically symmetric behavior of atomic dipolar gases. The directional dependence of the superfluid response suggests a pathway to realizing molecular quantum sensors with tunable anisotropy, enabling enhanced sensitivity to external perturbations in specific directions. This opens up exciting possibilities for gyroscopic or inertial sensing applications.

To investigate roton physics and the onset of supersolidity, we developed a quasi-one-dimensional theory for the excitation spectrum in the presence of transverse confinement. This approach allows us to track the emergence of roton minima in finite geometries. We find that increasing the ellipticity of the microwave polarization can reliably trigger roton softening and the spontaneous breaking of translational symmetry. These results are consistent with recent experimental observations in systems employing double-microwave shielding \cite{zhang2025observation}.

Interestingly, our results suggest that the order of the transition from a superfluid to a supersolid phase depends sensitively on the ellipticity parameter. At small ellipticity, the transition appears sharp and possibly first-order, while at larger ellipticity, it becomes continuous. This behavior implies that the nature of the phase transition can be tuned by adjusting the polarization of the microwave shielding. From an experimental standpoint, accessing the supersolid phase via a continuous transition is advantageous, as it minimizes heating and preserves coherence across the system---conditions that are essential for the robust preparation of molecular supersolids.

Finally, we note that current experiments using single-microwave shielding are often limited by significant three-body recombination losses or inelastic two-body collisions near degeneracy. Our analysis identifies a parameter regime where these loss mechanisms are minimized, and which remains accessible with current experimental capabilities. To mitigate these losses, recent experiments have deployed double-microwave shielding, which effectively removes the $\xi$-independent terms of the long-range interaction $V_3$, resulting in a weaker effective dipole-dipole interaction. Our results strongly indicate that similar roton-driven crystallization phenomena can be realized in the double-shielded case using the same theoretical machinery. A detailed analysis of this regime will be presented in a forthcoming work, and other contemporary avenues such as static electric field shielding could also be investigated~\cite{mukherjee2025effective}.

\begin{acknowledgments}
We thank Andreas Schindewolf for very helpful comments on the manuscript. We are grateful to Philipp St\"urmer, Malte Schubert, Koushik Mukherjee, Lila Chergui, and Deepak Gaur for discussions. This work was financially supported by the Knut and Alice Wallenberg Foundation (Grants No. KAW 2018.0217 and KAW 2023.0322) and the Swedish Research Council (Grant No. 2022-03654VR).
\end{acknowledgments}

\appendix

\section{Effective potential under elliptical microwave shielding}\label{app:effpot}
\begin{figure}
    \centering
    \includegraphics[width=4.4cm]{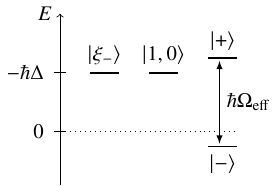}
    \caption{Energy level structure for microwave shielded molecules in the rotating frame.}
    \label{fig:levels}
\end{figure}

In this Appendix we derive the potential $V_\mathrm{SMS}(\bm{r})$ and show that previously suggested constraints on $\xi$~\cite{deng2023effective} are not needed. We start by considering a single molecule, which may occupy the rotational states $\ket{J,M}$. We introduce a radiation that only couples the state $\ket{0,0}$ with $\ket{1,M}$ states, hence we restrict the Hilbert space to the four-dimensional subspace corresponding to $J=0$ and 1. The corresponding non-interacting Hamiltonian reads $\hat H_0=\hbar\omega_\text{eg}\sum_{M=-1}^1\ket{1,M}\bra{1,M}$, where $\hbar\omega_\text{eg}=2B_\mathrm{rot}$ and $B_\mathrm{rot}$ is the rotational constant. A microwave field of frequency $\omega$ and amplitude $E_0$ can be expressed as
\begin{align}
    \bm{E}(t) = -\frac{E_0}{2}e^{-i\omega t}\left[\cos\xi\hat{\bm{e}}_+ +\sin\xi \hat{\bm{e}}_-\right] + c.c. ,
\end{align}
where $\hat{\bm{e}}_\pm=\mp (\hat{\bm x} \pm\hat{\bm y})/\sqrt{2}$ are circular unit vectors. The molecule-field interaction in the dipole approximation reads $\hat H_1 = -\hat{\bm{d}}\cdot\bm{E}(t) $, where the dipole operator is 
\begin{align}
    \bm{\hat d} = \frac{d}{\sqrt 3}\big(&\bm{e}_z\ket{0,0}\bra{1,0} + \bm{e}_+\ket{0,0}\bra{1,1} \nonumber\\&+\bm{e}_-\ket{g}\bra{1,-1}\big) + h.c.
\end{align}
Applying the unitary transformation $\hat U(t)=\exp{i\omega(\ket{1,0}\bra{1,0}+\ket{1,1}\bra{1,1}+\ket{1,-1}\bra{1,-1})t}$ and the rotating wave approximation, the total Hamiltonian $\hat H=\hat U(\hat H_0+\hat H_1)\hat U^\dagger-i\hbar \hat U\dot{\hat{U}}^\dagger$ takes the form
\begin{align}
    \hat H = &-\hbar\Delta\left(\ket{1,0}\bra{1,0}+\ket{\xi_+}\bra{\xi_+}+\ket{\xi_-}\bra{\xi_-}\right)\notag \\
    &+\frac{\hbar\Omega}{2}\left(\ket{\xi_+}\bra{0,0}+\ket{0,0}\bra{\xi_+}\right) ,
\end{align}
where we introduced the state vectors $\ket{\xi_+}=\cos\xi\ket{1,1} +\sin\xi \ket{1,-1}$ and $\ket{\xi_-}=-\sin\xi\ket{1,1}+\cos\xi\ket{1,-1}$, as well as the Rabi frequency $\Omega=dE_0/\sqrt 3$ and detuning $\Delta=\omega-\omega_\text{eg}$. The above Hamiltonian has eigenstates $\ket{+}=u\ket{0,0}+v\ket{\xi_+}$, $\ket{-}=u\ket{\xi_+}-v\ket{0,0}$, $\ket{\xi_-}$ and $\ket{1,0}$. The corresponding energies are $E_\pm=\hbar(-\Delta\pm\Omega_\text{eff})/2$ and $E_{\xi_+}=E_{\xi_-}=-\hbar\Delta$, where the effective Rabi frequency is $\Omega_\text{eff}=\sqrt{\Omega^2+\Delta^2}$. The amplitudes $u$ and $v$ are given by $u=\sqrt{(1+ \Delta/\Omega_\text{eff})/2}$ and $v=\sqrt{(1- \Delta/\Omega_\text{eff})/2}$. A schematic representation of the energy levels in given in Fig.~\ref{fig:levels}. In this work, we assume $\Delta<0$ and consider molecules in the eigenstate of largest energy $\ket{+}$.

The interaction between molecules is given by the operator
\begin{align}
    \hat V = \frac{1}{4\pi\epsilon_0r^3}\left[\bm{\hat d}_1'\cdot\bm{\hat d}_2' -3(\bm{\hat d}_1'\cdot\bm{\hat r})(\bm{\hat d}_2'\cdot\bm{\hat r})\right],
\end{align}
where the dipole operator for molecule $i$ is given by $\hat{\bm{d}}_i'=\hat U \hat{\bm d}_i \hat U^\dagger$ in the rotating frame. Using perturbation theory, we calculate the energy shift of the unperturbed two-molecule state $\ket{++} = \ket{+}\otimes\ket{+}$ induced by the dipole-dipole interaction $\hat V$, to which we apply the rotating wave approximation. Within the Born-Oppenheimer approximation, this corresponds to the effective potential $V_\text{SMS}(\bm r)$. The first order term is given by
\begin{align}
    V^{(1)} = \bra{++}\hat V\ket{++}.
\end{align}
Carrying out the calculation one simply gets $V^{(1)}=V_3$. 
The second order term reads
\begin{align}\label{eqn:v2}
    V^{(2)} = \sum_{(\nu_1,\nu_2)\neq(+,+)}\frac{|\bra{++}\hat V\ket{\nu_1\nu_2}|^2}{2E_+-E_{\nu_1}-E_{\nu_2}}
\end{align}
where the sum runs on all the two-molecule states $\ket{\nu_1,\nu_2}$ other than $\ket{++}$. It is useful now to change to the basis of symmetrized $\ket{\nu_1\nu_2}_s=(\ket{\nu_1\nu_2}+\ket{\nu_2\nu_1})/\sqrt 2$ and antisymmetrized states and note that $\bra{++}\hat V\ket{\nu_1\nu_2}_a=\bra{++}\hat V\ket{\xi_-\xi_-}=\bra{++}\hat V\ket{00}=0$. Before proceeding, we note that each term of the sum in Eq.~(\ref{eqn:v2}) is inversely proportional to the difference between the energy of the state $\ket{++}=2E_+$ and that of $\ket{\nu_1\nu_2}$. Accordingly, the contributing states are $\ket{+\xi_-}_s$, $\ket{+0}_s$ and $\ket{\xi_-0}_s$. Moreover, one has $\bra{++}\hat V\ket{\xi_-0}_s=0$. Therefore, it is only necessary to consider the following matrix elements:
\begin{align}\label{eq:mel}
    |\bra{++}\hat V\ket{+\xi_-}_s|^2 = &\frac{d^4u^4v^2}{2(4\pi\epsilon_0)^2r^6}\sin^4\theta\nonumber \\
    &\times\left(\cos^2 2\xi\cos^2 2\phi + \sin^2 2\phi\right) \\ 
    |\bra{++}\hat V\ket{+0}_s|^2 = &-\frac{d^4u^4v^2}{4(4\pi\epsilon_0)^2r^6}\sin^2 2\theta\nonumber\\&\times\left(\cos 2\phi \sin 2\xi -1\right).
    \label{eq:mel2}
\end{align}

After a lengthy, but straightforward calculation substituting Eqs.~\eqref{eq:mel} and~\eqref{eq:mel2} into Eq.~\eqref{eqn:v2}, one gets $V^{(2)}= V_6$. Finally, the effective interaction between molecules is given by $V_\text{SMS}(\bm r)=V^{(1)}+V^{(2)}\equiv V_3+V_6$ with $V_6$ and $V_3$ as given in Eq.~(\ref{eqn:srpotential}) and Eq.~(\ref{eqn:lrpotential}) without any restriction or approximation involving $\xi$.

\section{Evaluation of $\mathcal{K}_5$}\label{app:K5}

\begin{figure}
    \centering
    \includegraphics[width=1\linewidth]{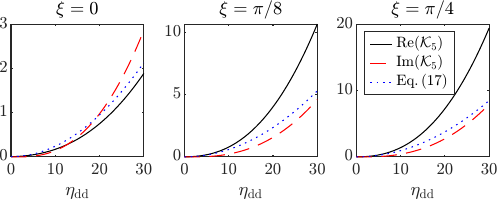}
    \caption{Real and imaginary parts of the $\mathcal{K}_5$ function over a range of $\edd$. The amplitude of each curve is reduced by a factor of 1000 for visibility.}
    \label{fig:K5}
\end{figure}

As discussed in the main text, the beyond-mean-field energy contribution relies on a calculation of the integral given in Eq.~\eqref{eqn:K5}. A small $\edd$ expansion yields the approximation Eq.~\eqref{eqn:K5app}, which trivially reduces to the known antidipolar (dipolar) limit for $\xi = 0^\circ$ ($\xi = 45^\circ$). However, in this work we primarily explore the regime $\edd \gg 1$, where this expansion is no longer valid. Therefore, all simulations are performed using the full integral formulation of $\mathcal{K}_5$. In Fig.~\ref{fig:K5}, we show the real and imaginary parts of $\mathcal{K}_5$ as a function of $\edd$ and $\xi$, with the second-order approximation overlaid for comparison. At large $\edd$ and $\xi=0^\circ$, the imaginary component dominates over the real one, and the observed agreement between our two methods is therefore particularly striking.

\section{Numerical details}\label{app:V3k}

The PIMC results in this work are obtained by averaging over $10^{4}-10^{5}$ Monte Carlo sweeps, with each sweep consisting of $\mathcal O(N)$ updates involving single polymers or entire permutation cycles \cite{spada2022}. In Sec.~\ref{sec:super}, we use a number of beads $M$ between 1000 and 8000, in order to eliminate finite-$M$ effects. In Sec.~\ref{sec:SS}, $M$ has been set to 2100.

To solve the eGPE-M, Eq.~\eqref{eqn:gpe}, we employ a split-step Fourier method with grid sizes \((N_x,N_y,N_z) = (64, 128, 128)\) in a box of lengths \(\bm{L} = (40, 40, 60)\,a_\perp\), and an imaginary time-step \(\Delta t = -i\times10^{-4}\,\omega_\perp^{-1}\). We compare solutions obtained from multiple initial conditions (Gaussian, density-modulated with or without a central droplet, and uniform random noise) to identify the lowest energy state for Figs.~\ref{fig:SS} and~\ref{fig:contrast}. Additionally, in these figures we verify all modulated states, contrasts, and transition points using simulations with double the number of grid points in each direction. The example densities shown are taken from these fine-grained simulations.

The dipolar contribution to the eGPE-M is efficiently evaluated in momentum space using the convolution theorem
\begin{align}
    \int \mathrm{d}^3 \bm{r}'\, V(\bm{r} - \bm{r}') |\Psi(\bm{r}')|^2 
    = \mathcal{F}^{-1}\left[ \tilde{V}(\bm{k})\, \tilde n(\bm{k}) \right]\,,
\end{align}
where $\mathcal{F}^{-1}$ denotes the inverse Fourier transform, $\tilde n(\bm{k})$ is the Fourier transform of the density, and the interaction potential in momentum space is defined in Eq.~\eqref{eqn:DDIk}.

In practice, evaluating this expression using fast Fourier transform (FFT) methods introduces numerical artifacts due to the implicit periodic boundary conditions. Since the dipole-dipole interaction is long-ranged, these artificial periodic images of the condensate interact with each other, leading to unphysical contributions in the computed potential.

To mitigate this issue, a spherical cut-off is applied to the DDI in real space, suppressing contributions beyond a radius $R_c$ that exceeds the size of the physical system but remains smaller than the simulation box. This corresponds to calculating the Fourier transform of the truncated potential $V_3(\bm{r})\,\Theta(R_c - |\bm{r}|)$, where $\Theta$ is the Heaviside step function. The resulting expression for the Fourier-transformed dipolar potential becomes
\begin{align}
    \tilde{V}^{R_c}_3(\bm{k}) = \tilde{V}_3(\bm{k}) \left[ 
    1 + 3\frac{\cos(R_c k)}{R_c^2 k^2} 
      - 3\frac{\sin(R_c k)}{R_c^3 k^3} 
    \right]\,,
\end{align}
which is identical to the result obtained for dipolar atoms~\cite{ronen2006bogoliubov}. This correction effectively removes contributions from periodic images while preserving the physical long-range behavior within the region of interest.

\bibliography{refs.bib}

\end{document}